\def\year{2015}
\documentclass[letterpaper]{article}

\usepackage{aaai}
\usepackage{courier}
\usepackage{booktabs}
\usepackage{url}
\usepackage{array}
\usepackage{amsmath}

\usepackage{latexsym}
\usepackage{color}
\usepackage[usenames]{xcolor}
\usepackage{amssymb}
\usepackage{rotating}
\usepackage{multirow}
\usepackage{pdfsync}
\usepackage{url}
\usepackage{enumitem}

 \newif\ifdraft
\drafttrue

\begin{document}
%
\title{Quantifying Public Response towards Islam on Twitter after Paris Attacks}
\author{Walid Magdy$^1$, Kareem Darwish$^1$, and Norah Abokhodair$^{1,2}$\\
$^1$Qatar Computing Research Institute, HBKU, Doha, Qatar\\
$^2$Information School, University of Washington, Seattle, USA\\
Email: \{wmagdy,kdarwish\}@qf.org.qa, noraha@uw.edu\\
Twitter: @walid\_magdy, @kareem2darwish, @norahak\\
}
\maketitle


\begin{abstract}
\begin{quote}
The Paris terrorist attacks occurred on November 13, 2015 prompted a massive response on social media including Twitter, with millions of posted tweets in the first few hours after the attacks. Most of the tweets were condemning the attacks and showing support to Parisians. One of the trending debates related to the attacks concerned possible association between terrorism and Islam and Muslims in general. This created a global discussion between those attacking and those defending Islam and Muslims. In this paper, we provide quantitative and qualitative analysis of data collection we streamed from Twitter starting 7 hours after the Paris attacks and for 50 subsequent hours that are related to blaming Islam and Muslims and to defending them. We collected a set of 8.36 million tweets in this epoch consisting of tweets in many different of languages. We could identify a subset consisting of 900K tweets relating to Islam and Muslims. Using sampling methods and crowd-sourcing annotation, we managed to estimate the public response of these tweets. Our findings show that the majority of the tweets were in fact defending Muslims and absolving them from responsibility for the attacks. However, a considerable number of tweets were blaming Muslims, with most of these tweets coming from western countries such as the Netherlands, France, and the US.  
%
%
%
%
\end{quote}
\end{abstract}



\section{Introduction}
\label{sec:intro}
The coordinated terrorist attacks on multiple spots in Paris on Friday November 13, 2015 prompted a massive worldwide response on social media including Twitter.  The response was mostly focused on expressing outrage at the attacks and sympathy for the victims.  After the Islamic State in Iraq and Syria (ISIS) announced their responsibility the the attacks~\cite{ISISclaims}, reactions started to appear on social media blaming Muslims for the attacks and linking terrorism to Islam. This reaction resulted in a counter movement by people from around the world to disassociate Muslims from the perpetrated crimes and clarify that ISIS does not represent Islam.  In this paper, we present the results of our preliminary quantitative data analysis on the data we collected from the incident to highlight both reactions and show the distribution of each reaction across different regions in the world.
%
%
%

After 7 hours of the terrorist attacks in Paris, We started collecting tweets on the topic. We collected tweets that contained words and hashtags that were relevant to the event.
We managed to collect a set of 8.36 million tweets in the 50 hours following the attacks. Out of those tweets we could identify a set of more than 900,000 tweets talking about Muslims and Islam in different languages. We used this set of tweets to analyze the reaction towards Muslims after the attacks. We applied sampling, manual labeling, and label propagation to estimate the attitude distribution. We estimated the distribution across the top different languages and over countries.

Our analysis shows that the majority of tweets showing attitude towards Muslims and Islam after the Paris attacks were positive towards Muslims: defending Muslims and disassociating them from the attacks. However, there was still a considerable number of tweets attacking Muslim and relating terrorism to Islam. Negative attitudes ranged from asking governments to take action against Muslims to calls for exterminating all Muslims as in \#KillAllMuslims.  

The work presented in this paper is just a preliminary quantitative and qualitative study about the direct public response after the Paris terrorist attacks towards Islam and Muslims. A deeper study is to be held on the background of the Twitter users who showed positive or negative attitudes towards Muslims after the attacks in hopes of understanding the motivations for their reactions.

The rest of the paper is organized as follows: Section 2 gives some background on Paris attacks and reports some related work on Islamophobia. Section 3 describes the data collection, including the crawling process, sampling methodology, and annotation process. Section 4 presents the attitude distribution over languages and regions, and gives some qualitative analysis to the top trending tweets. Finally, section 5 concludes the paper and defines the potential future directions. 



\section{Background}
\label{sec:relatedwork}

\subsection{The Terrorist Attacks on Paris 2015}
On the evening of 13 November 2015, several terrorist attacks occurred simultaneously in Paris, France. At 20:20 GMT, three suicide bombers struck near the Stadium where a football match between France and German was being played. Other suicide bombings and mass shootings occurred a few minutes later at cafes, restaurants and a music venue in Paris~\cite{AttacksTimeline,WhatHappened}.

More than 130 people were killed in those attacks and 368 were injuries, including 80 to 99 people seriously injured. These attacks are considered the deadliest in France since World War II~\cite{Indepedent}.  The Islamic State of Iraq and Syria (ISIS)\footnote{also known as Islamic State of Iraq and the Levant (ISIL)} claimed the responsibility for those attacks~\cite{ISISclaims} as a response to the French airstrikes on the its targets in Syria and Iraq.

\subsection{Anti-Muslim rhetoric}
Some papers in the literature referring to anti-Muslim actions
as Islamophobia.  There is still a debate on the exact meaning and characteristics of Islamophobia in the literature. Some scholars regard it as a type of hate speech and others as a type of racism~\cite{awan2014islamophobia}.  In most cases it refers to the phenomenon of negatively representing Muslims and Islam in Western media. 

According to the Oxford English Dictionary, the word means ``Intense dislike or fear of Islam, esp. as a political force; hostility or prejudice towards Muslims.'' The term Islamophobia was coined in 1997 by the Runnymede Trust report after organizing a commission to address the manifestation of anti-Muslim and anti-Islam sentiment in British media. In the Runnymede report, Islamophobia was defined as ``an outlook or world-view involving an unfounded dread and dislike of Muslims, which results in practices of exclusion and discrimination.''~\cite{runnymede1997islamophobia}. The goal of coining the term was to recognize the issue and to condemn the negative emotions directed at Islam or Muslims~\cite{runnymede1997islamophobia}. 


Online Islamophobia remains under researched~\cite{copsey2013anti}. Based on the earlier provided definitions, we start exploring the manifestation of Islamophobia on social media. In this study we focused on the expression of Islamophobia on Twitter and its characteristics with examples from the data. It is important to note that we are not talking here about legitimate critical analysis of Muslim politics, society and even culture, which we consider healthy to debate. We follow the characteristics and types mentioned in the definitions. In our data we use the same measure and criteria that is used to identify hate-speech, online abuse and cyber-bulling in online environments. 

Awan~\shortcite{awan2014islamophobia} defines online Islamophobia, as ``Anti- Muslim hate is prejudice that targets a victim in order to provoke, cause hostility and promote intolerance through means of harassment, stalking, abuse incitement, threatening behavior, bullying and intimidation of the person or persons, via all platforms of social media''. Similar to \cite{awan2014islamophobia,runnymede1997islamophobia} we recognize the need to provide a definition of this phenomenon to so it can be identified and further discussed by policy makers and technology designers. 

In this paper, we focus more on quantifying the amount of attack and defense to Islam and Muslims, and we keep the motivation for these attitudes and if it is related to Islamophobia for future work.

\section{Data Collection}


\subsection{Streaming Tweets on the Attacks}
\label{sec:data}
In few hours after the attacks, the trending topics on Twitter were mostly referring to the Paris attacks and expressing sympathy for the victims. We used these trending topics to formulate a set of terms for streaming tweets using the Twitter REST API. We also used general terms referring to terrorism and Islam, which were hot topics at that time. We continuously collected tweets between 5:26 AM (GMT) (roughly 7 hours after the attacks) on November 14 and 7:13 AM (GMT) on November 16 (approximately 50 hours in total).
The terms we used for collecting our tweets were: \textsf{Paris}, \textsf{France}, \textsf{PorteOuverte}, \textsf{ParisAttacks}, \textsf{AttaquesParis}, \textsf{pray4Paris}, \textsf{paryers4Paris}, \textsf{terrorist}, \textsf{terrorism}, \textsf{terrorists}, \textsf{Muslims}, \textsf{Islam}, \textsf{Muslim}, \textsf{Islamic}.  In total we collected 8.36 million tweets. 
Since we were using the public APIs and due to Twitter limits, we were sampling the Twitter stream and our samples were reaching the set limits. However, since we were searching using focused keywords, streaming most probably captured a large percentage (if not the majority) of on-topic tweets.  In all, we were collecting 140k to 175k tweets per hour.  Post our collection, we checked the counts of the terms we used for search in Topsy\footnote{\url{http://topsy.com/}}, and our estimates indicate that the number of tweets on Twitter that matched our search terms was slightly more than 12 million tweets. Also, since we were using mostly English words/hashtags and a few French ones, then our expectations is that were collecting mostly English, to a lesser extent French tweets, and even less tweets from other languages. However, the main term, \textsf{Paris}, is language independent for most Latin languages, which resolves part of this problem.



\begin{figure}
\centering
\includegraphics[width=\columnwidth]{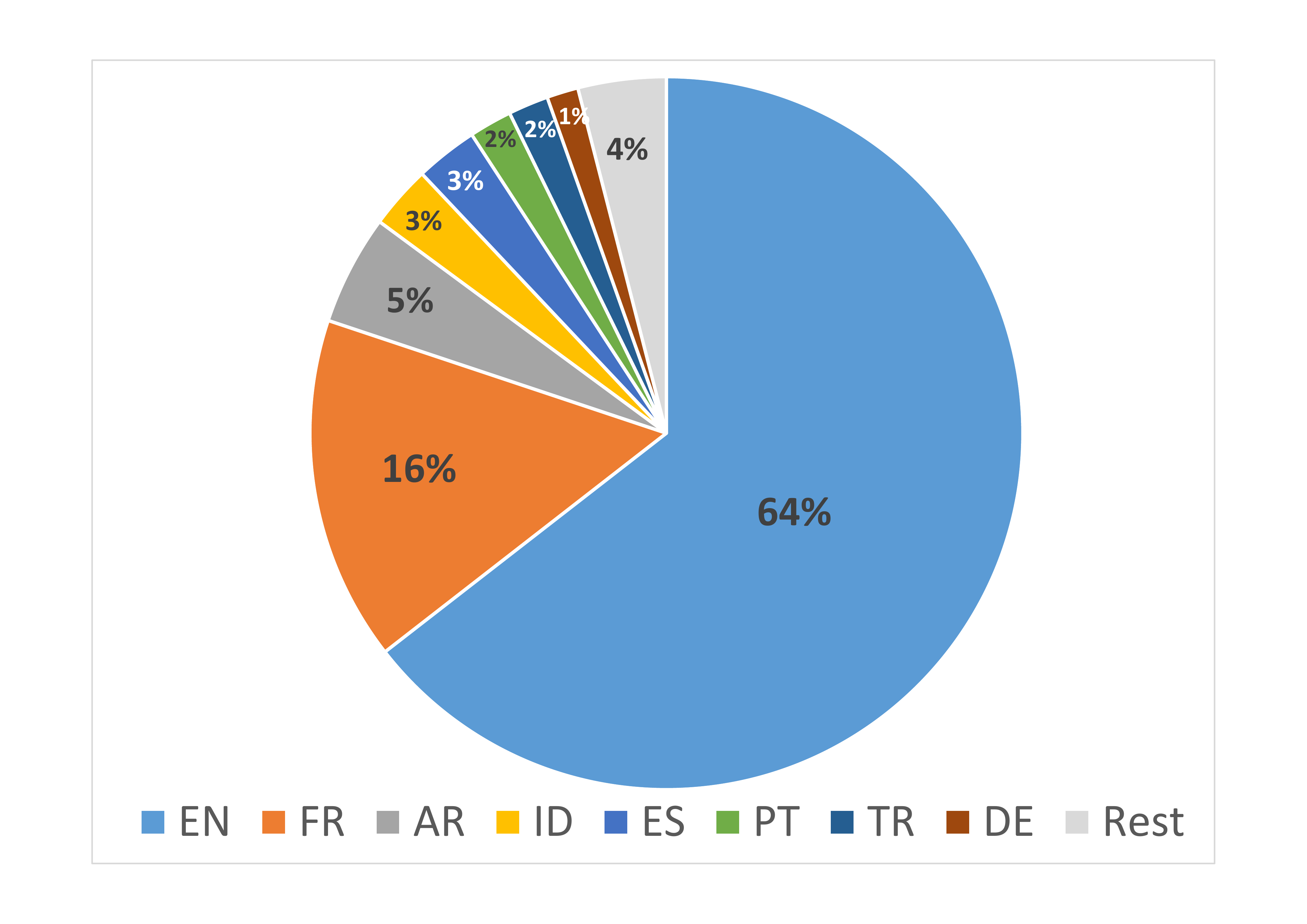}
\caption{\label{LangStats}Language distribution of the tweet collection}
\end{figure}

A language identification process was then applied to each of the tweets to understand the distributions of languages in our collection. We used an open source language detection java-based library\footnote{\url{https://github.com/shuyo/language-detection}} to detect the language of tweets. Figure~\ref{LangStats} shows the language distribution of our tweet collection. As shown, the majority of the tweets (64\%) are in English, which is expected since English is the largest language on Twitter and people tend to show comment on a global event like this in the language that is mostly known. The second language was French, the language used at the location of the attacks. Surprisingly, the third language was Arabic, though all of the keywords used for crawling were in Latin letters. However, it was clear that Arabs were commenting on the topic in their own language and adding English hashtags to make their tweets discoverable.  In all, there were tweets in 52 different languages with varying volumes.  We decided to restrict our analysis to the 10 languages with the most number of tweets.  These languages were: English, French, Spanish, German, Dutch, Italian, Portuguese, Arabic, Turkish, and Indonesian.  We further excluded language that are predominantly spoken in majority Muslim countries, namely: Arabic, Turkish, and Indonesian.  We dropped these languages from our analysis, since we were more interested in the public response to the attacks in non-Muslim countries.

\subsection{Identifying Tweets on Islam}
\label{sec:islamtweets}
For identifying the tweets about Islam and Muslims,  we filtered the tweets using terms that refer to Islam, such as Islam, Muslim, Muslims, Islamic, and Islamist. Since we were interested in the 7 languages of interest in the collection, we translated the filter terms into these languages, and search for all tweets matching these terms.

Out of the 8.36 million tweets, we could extract 912,694 tweets mentioning something about Islam in one of the 7 languages. This number constitutes 11\% of the collected tweets, which is considered significant percentage of the tweets of the topic.



\section{Identifying Attitudes towards Muslims}
\label{sec:approach}
\subsection{Sampling Tweets}
The number of tweets pertaining to Islam and Muslims was too large to be fully manually annotated. In order to find out the attitude of the tweets, we sampled the data collection by getting a representative sample of tweets for each language. We used a sample size calculator~\footnote{\url{http://www.surveysystem.com/sscalc.htm}} to calculate the sample size of each language that would lead to an estimation of the attitude distribution with error less that $\pm$2.5\% (confidence interval = 2.5\%) and a confidence level of 95\%. We then selected a set of tweets at random from each language based on the sample size to be manually annotated. Table~\ref{tab:samples} shows the number of tweets in each of the seven western languages and the samples extracted from each. As shown, the sample size is not linear to the full size, since these sizes is selected to maintain a given confidence and quality.


\begin{table}[t]
\begin{center}
{
\begin{tabular}{crr}
\hline
Language & Size & Sample\\\hline
EN & 753,476 & 1,534\\
FR & 63,410 & 1,500\\
ES & 15,726 & 1,400\\
DE & 6,388 & 1,239\\
NL & 4,406 & 1,139\\
IT & 3,825 & 1,096\\
PT & 2,194 & 904\\
\hline
\end{tabular}
}
\end{center}
\caption{\label{tab:samples}Number of tweets about Islam of the top seven western languages, and the sample size of each}
\end{table}

\subsection{Tweets Annotation}
The sets of sampled tweets were then submitted to crowdflower~\footnote{\url{http://www.crowdflower.com/}} to be manually annotated. We asked annotators to label each of the tweets with one of three labels:
\begin{itemize}
\item \textbf{Defending}: the tweet is defending Islam and/or Muslims against any association to the attacks.
\item \textbf{Attacking}: the tweet is attacking Islam and/or Muslims as being responsible for the terrorist attacks.
\item \textbf{Neutral}: the tweet is reporting news, not related to the event, or talking about ISIS in specific and not Muslims in general.
\end{itemize}

We submitted each set of tweets in one language in a separate annotation job on crowdflower, and restricted the annotators to speakers of the language of the tweets being annotated. Each tweet was annotated by at least 3 annotators, and the majority voting is taken for selecting the final label. In crowdflower, a golden control set of tens of tweets is required to be used for assessing the quality of work of annotators, where low quality ones are discarded. To prepare the golden set of tweets, we had to translate some of the non-English tweets into English using Google translate, and we used the ones with clear translations in the golden set.

\subsection{Location Identification}
Sampling the tweets by language gives a good estimation to the attitude across languages. However, finding the attitude by language does not give an accurate indication on the attitude by country, since one languages could be spoken in multiple countries, and one country may have more than one spoken language. For instance, English tweets were generated from many countries around the world.

The total number of annotated sample tweets were 8,716 tweets. Applying location analysis based on this number only would be very sparse, since there are large number of locations, and it would be difficult to get good coverage of different locations in this sample. However, we noticed that some of the tweets in our collection are actually retweets or duplicates of other tweets. Thus, we applied label propagation to label the tweets in our collection that have identical text to the labeled tweets. To detect duplicates and retweets, we normalized the text of the tweets by applying case folding, and filtering out URLs, punctuation, and name mentions. Tweets in the collection that matched the annotated sample tweets after text normalization were then labeled with the same label. This label propagation process led to the labeling of 336,294 of the tweets in the collection, which represents over than 36\% of the tweets referring to Islam.

For the 336k labeled tweets, we extracted the user declared locations for mapping to countries. The location field in Twitter is optional, so users can leave it blank. In addition, it is free text, which means that there is no standard method for writing the location. This renders a large portion of these locations as unusable.  Example locations are ``in the heart of my mom'',``the 3rd rock from the son'', and ``at my house''. This is a common problem in social media in general and in Twitter in particular, as demonstrated in~\cite{hecht2011tweets}.

In our work, we used a semi-supervised method for mapping the locations to countries. Our method applies the following steps:
\begin{enumerate}
\item A list of countries of the world and their most popular cities were collected from Wikipedia and saved in a database, where each city is mapped to its correspondent country.
\item The list of states of the united states and their shortcuts along with the top cities in these states were added to the database.
\item Location strings are normalized by case folding and removing accents from letters. For example, ``M\'{e}xico'' will be normalized to ``mecixo''.
\item If the location string contains a country name, it is mapped to the country. Otherwise, the string is searched for having a city/state name in our database, then it is mapped to its corresponding country. In case multiple countries/cities exist in the location string, the mapping is applied to the first one in the location string.
\item All unmapped locations appearing at least 10 times are then manually mapped to corresponding countries whenever possible (since sometimes junk locations occurs frequently, such as ``earth''). All newly mapped locations are then added to the database, then an additional iteration of matching in the previous step is applied to map location strings containing any of the newly added locations.
\end{enumerate}

Initial application of our approach on the 336k locations, we found that 125,583 of these location were blank. In addition, 41,905 were locations of tweets labeled as ``neutral'', which were not of much interest in our analysis. The remaining non-blank locations of non-neutral tweets were 168,807 (with 76,894 unique locations). Using the above algorithm, we managed to map 106,411 locations (42,140 unique) to countries. 

\begin{figure}
\centering
\includegraphics[width=\columnwidth]{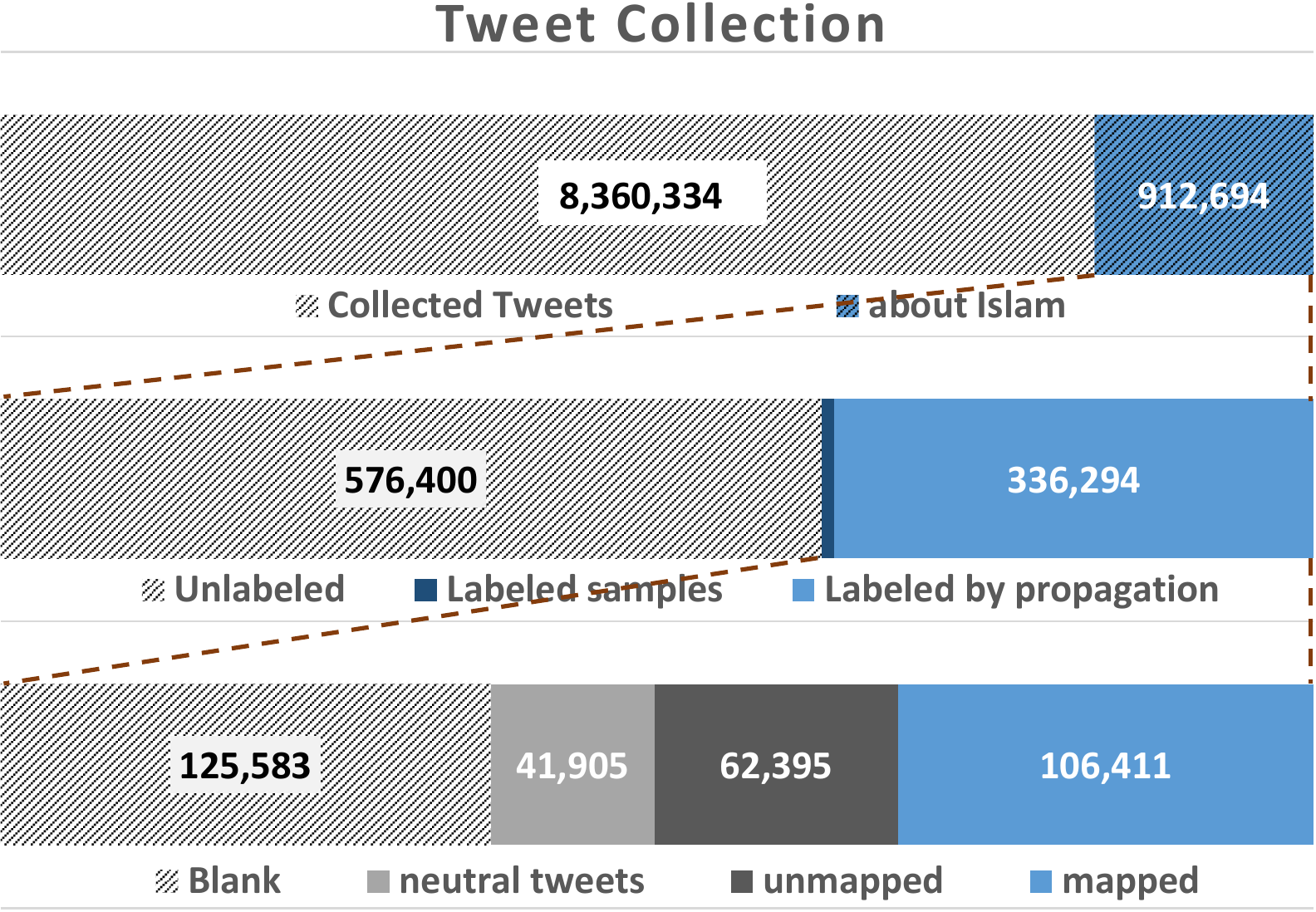}
\caption{\label{Collection}Summary of the tweet collection used in this study}
\end{figure}

Figure~\ref{Collection} summarizes the tweet collection in hands and all the steps applied to get the annotated data.

\section{Results}
\label{sec:results}

The tweet samples got annotated on crowdflower with an average inter-annotator agreement of 77.7\%, which is considered high for a three-choices annotation task annotated by at least three different annotators. The percentage of disagreement among annotators shows that some tweets are not straightforward to label. Usually this occurs between neutral and one of the other attitudes. In the following, we display the results we got from our analysis of the tweets.

\subsection{Distribution of Attitudes by Language}

\begin{figure*}[t]
\centering
\includegraphics[width=\textwidth]{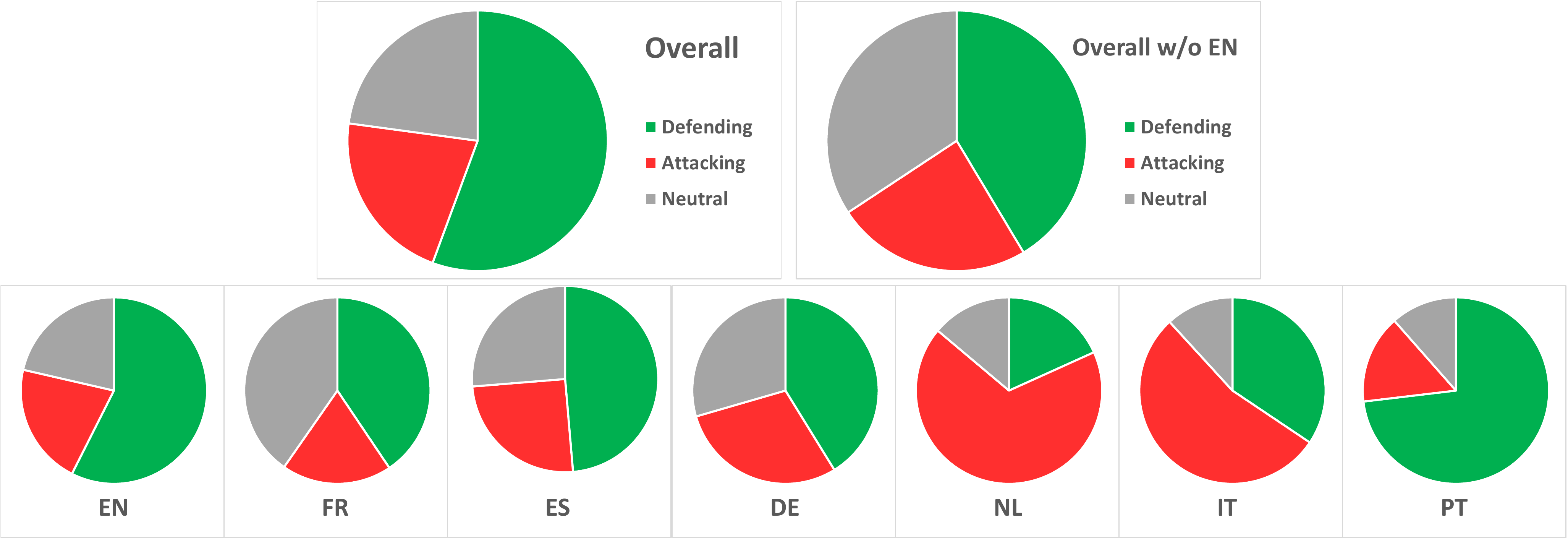}
\caption{\label{LangDist}Language distribution}
\end{figure*}

\begin{table*}
\begin{center}
{
\begin{tabular}{ll}
\hline
Positive & Negative\\\hline
\#MuslimsAreNotTerrorist (34,925) & \#IslamIsTheProblem (3,154)\\
\#MuslimAreNotTerrorist (17,759) & \#RadicalIslam (1,618)\\
\#NotInMyName (4,728) & \#StopIslam (1,598)\\
\#MuslimsStandWithParis (1,228) & \#BanIslam (460)\\
\#MuslimsAreNotTerrorists (1,106) & \#StopIslamicImmigration (333)\\
\#ThisisNotIslam (781) & \#IslamIsEvil (290)\\
\#NothingToDoWithIslam (619) & \#IslamAttacksParis (280)\\
\#ISISareNotMuslim (316) & \#ImpeachTheMuslim (215)\\
\#ExtremistsAreNotMuslim (306) & \#KillAllMuslims (206)\\
\#ISISisNotIslam (243) & \#DeportAllMuslims (186)\\
\hline
\end{tabular}
}
\end{center}
\caption{\label{tab:hashtags}Examples of the top hashtags that refer to positive and negative attitudes towards Muslims. Count of each hashtags is provided between brackets}
\end{table*}

Figure~\ref{LangStats} shows the distribution of attitudes towards Muslims according to each language, and the overall distribution of all languages, which is estimated based on the size of each language in the collection. As shown, most of the tweets are towards defending the position of Muslim and Islam by disassociating them from the attacks. This was the situation across most of the languages with some variations. Tweets in only two languages, namely Dutch (NL) and Italian (IT), had the majority of the attitudes attacking Muslims and relating the attacks to Islam in general. The support to Muslims in Portuguese (PT) was the highest among all languages.

The language which has the largest percentage of neutral tweets was French (FR), which might be expected, since France was the scene of the attacks and most likely people there were busy following the news and its updates more than anyone else. Many of these updates referred to Islamic State, which made it existing in our collection about ``Islam''.

The overall finding of this analysis, is that some voices appeared on Twitter trying to link the attacks by ISIS on Paris to Islam, representing 21.5\% of the tweets on topic. However, most tweets, namely 55.6\%, were defending Muslims and disassociating between terrorism and Islam.

Table~\ref{tab:hashtags} shows some examples of the most frequent hashtags in our collection that refer to positive and negative attitudes towards Muslims. As shown, the hashtags on the left side disassociate between ISIS/Terrorism and Islam, while those on the right side mainly focus on a call to ban Muslims from entering western countries, and some of them go as far as calling for the extermination of all Muslims (\#KillAllMuslims). The magnitude of positive hashtags is much larger than those with negative attitude.


\subsection{Attitudes by Country}

\begin{figure}[ht]
\centering
\includegraphics[width=\columnwidth]{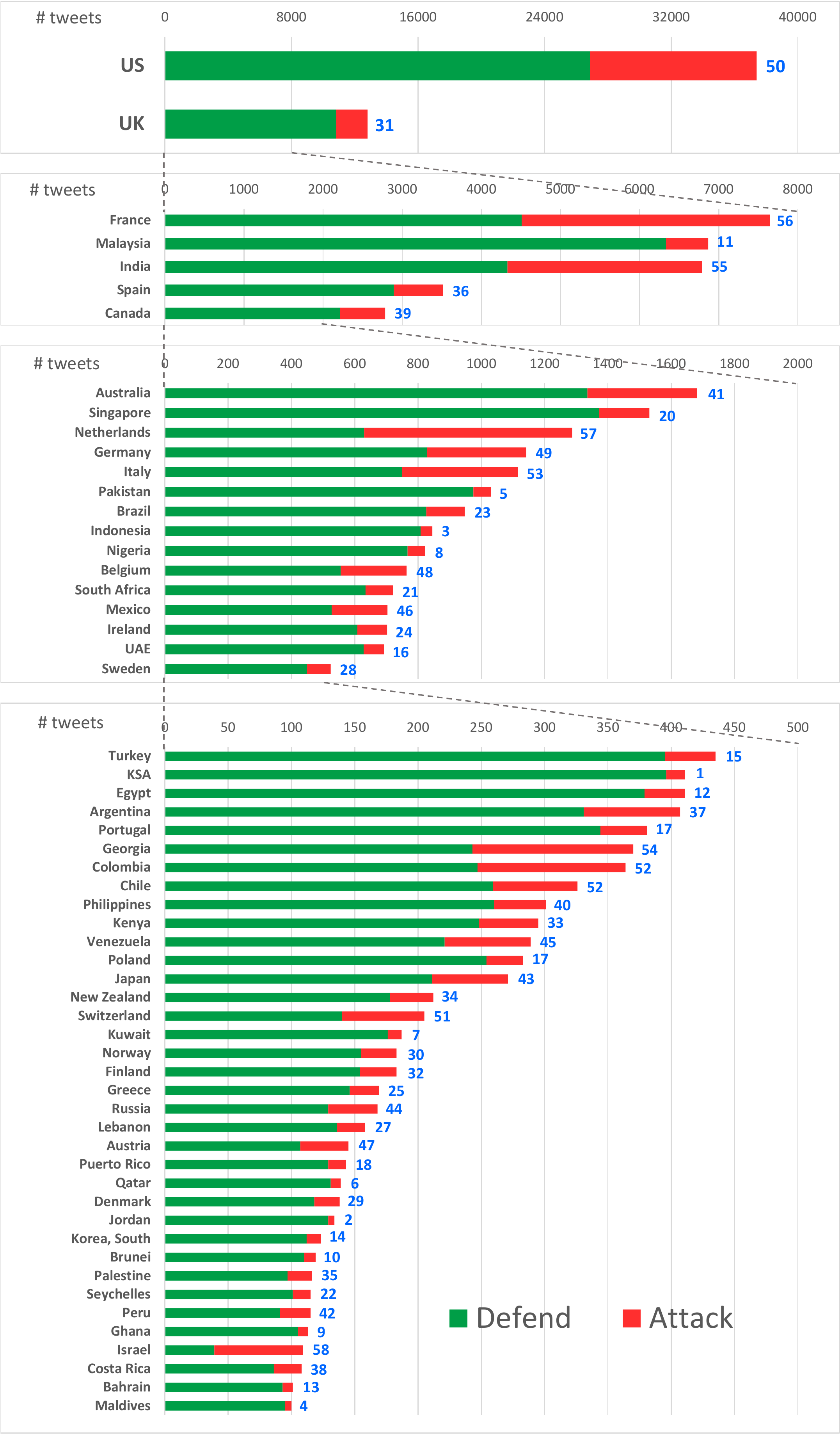}
\caption{\label{Locations}Attitude towards Muslims by country. Label beside each bar represents the rank of the country on defending Muslims.}
\end{figure}

As mentioned earlier, we managed to map the location of 106K tweets that have non-neutral attitude to countries. These locations were mapped to 144 different countries, which shows the global impact of the terrorist attacks that grabbed significant world wide attention.

Some of the countries had only few tweets mapped to them, which makes any observation for the global attitude of these countries far from being accurate. Thus, in our analysis, we only kept the countries which had at least 100 tweets mapped to it. Using this restriction, the number of countries reduced to 58 ones.

The united states (US) had the most number of tweets, namely 36.5\% of the mapped tweets, followed by the UK (12.5\%), France (7.5\%), Malaysia (6.7\%), India (6.6\%), and Spain (3.4\%). Each of the remaining countries had less than a 3\% share.

Figure~\ref{Locations} List the 58 countries that have more than 100 tweets mapped to them. For clarity, Figure~\ref{Locations} splits the graph into 4 pieces according to the order of magnitude of the number of tweets. The bars in the figure are split where the green and red parts represent positive and negative tweets towards Muslims respectively for each country. A rank for each country is displayed to the right of each bar according to the percentage of positive tweets\footnote{We ranked according to percentage defending Muslims, since it was the prevailing attitude}. As shown in Figure~\ref{Locations}, the countries with the highest percentages of positive tweets are mostly Muslim and/or Arabic countries, such as Saudi Arabia (KSA), Jordan, Indonesia, Maldives, Pakistan, and Qatar. Only two countries had more negative tweets than positive one, namely Israel and the Netherlands at ranks 58 and 57 respectively. They were followed by France, India, Georgia, and Italy at ranks 56, 55, 54, and 53 respectively. US, which is the country with the largest number of tweets, comes at the rank 50; while the UK, the country with the second highest number of tweets, comes at rank 31 with 85\% of its tweets defending Muslims.
 
Our analysis shows the large variations of attitudes between countries. As expected predominantly Muslim countries had the percentages of positive tweets. However, neighboring countries such as Spain (rank 36) and Italy (rank 53) had dramatically different percentages of positive/negative tweets.  This is also reflected in the percentage of Spanish and Italian language tweets, where roughly a quarter of Spanish tweets are negative compared to more than half of Italian tweets.  Similarly, the percentage of negative tweets is much higher in the Netherlands compared to Germany.  The large variation between juxtaposed countries is worthy of further study.  Further, the rank of the US is considerably low (rank 50).  This is likely due to partisanship related to the upcoming US presidential elections.  This is evident by hashtags such as \#ImpeachTheMuslim, see Table~\ref{tab:hashtags}, where when we explored tweets containing this hashtag, we found that the ``Muslim'' here is referring to President Barack Obama\footnote{Example tweet: \#ThanksObama \#isis \#paris \#ImpeachObama \#ImpeachTheMuslim \#RedNationRising \#NukeISIS \#StandwithTrump}
However, this requires further investigation. Figure~\ref{Locations} also shows some non-Muslim countries with very small Muslim populations and are ranked quite high, such as South Korea (rank 10) and Portugal (rank 17).
This also requires further investigation.


\subsection{Most Popular Tweets}
\begin{figure*}[ht!]
\centering
\includegraphics[width=\textwidth]{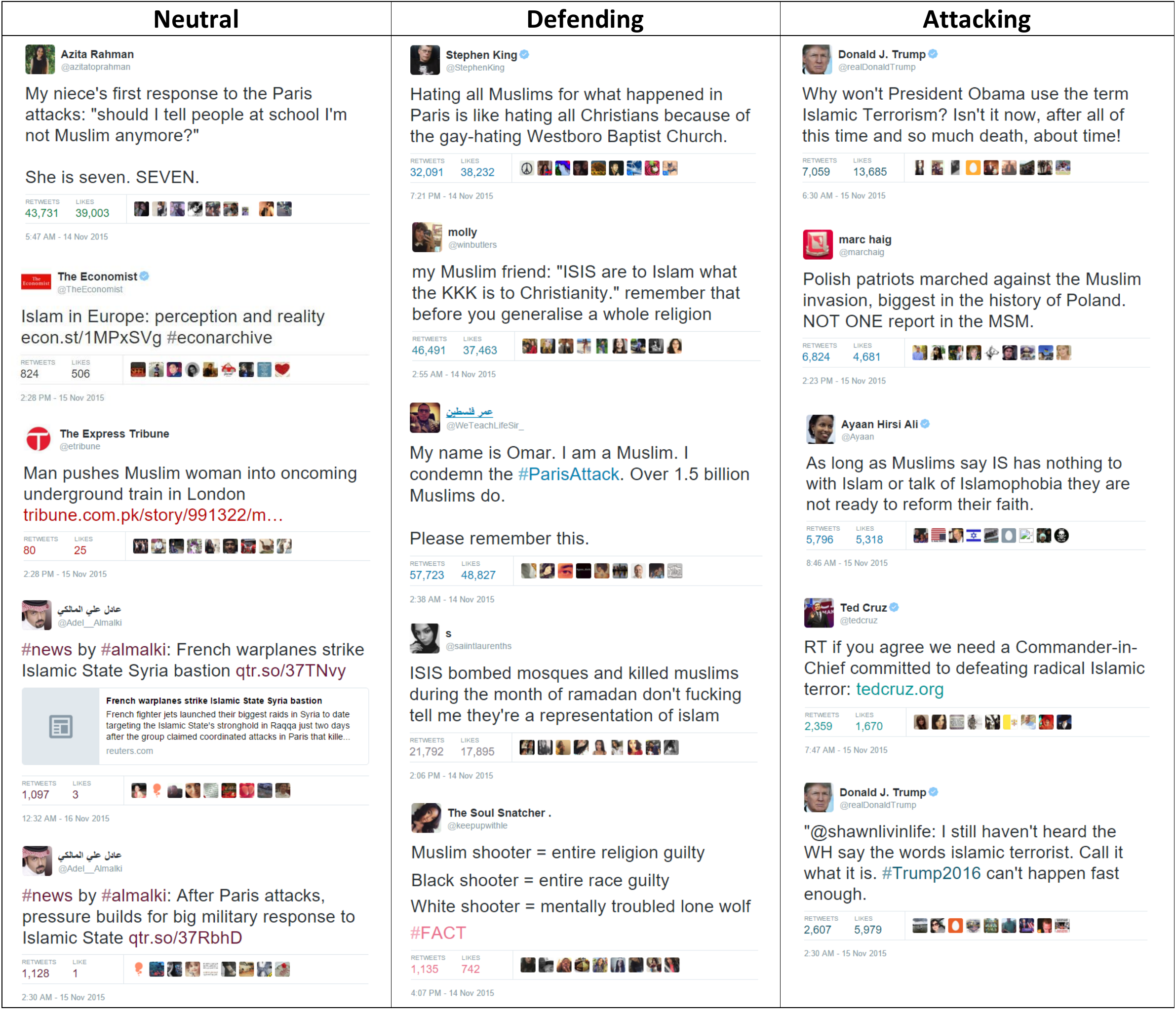}
\caption{\label{Examples}Most popular tweets for each attitude}
\end{figure*}

The label propagation step that we applied showed that large portion of the tweets in our collection are retweets. This refers to the presence of highly popular tweets that got retweeted thousands of times.
Our last research question was who are the most influential accounts in the discussion with positive or negative stances. In other words,  who was promoting the anti-Islam sentiment on Twitter in the time after the Paris attacks and who was against that sentiment. We decided to consider the 5 most retweeted tweets in each of the categories we identified earlier: neutral, defending, and attacking. Figure~\ref{Examples} illustrates the 5 most retweeted tweets with the account handle in each of the three categories (attacking, defending, and neutral). For the purpose of this paper we consider and discuss tweets in the list from the celebrity type accounts, someone who has both high content influence and high account influence.
Using both qualitative and quantitative analyses, we found that most of the interesting results appear in the `attacking' category.
However, we describe our observation on the three categories.

\subsubsection{Top Neutral and Positive Tweets}
The top 5 Neutral tweets were mostly about news, as expected. Only the top tweet, which received a large number of retweets (43,000+) was different. The tweet is coming from a seemingly Muslim female who has a moderate number of follower\footnote{1,826 followers by the time of writing the paper}. Her tweets was discussing the effect of the attacks on the Muslim community in US, where her young niece is afraid of telling her friends in school that she is Muslim. Although the tweet most probably would be retweeted by those disassociating Muslims from the attacks, the crowdflower annotators agreed that the tweet has no direct defense for Muslims, which is true to some extent.  The third tweet is concerned with a hate-crime that was perpetrated against a Muslim woman in London.

Regarding the most popular tweets defending the position of Muslims, two of them were tweeted by accounts apparently owned by Muslims. The top 2 tweets mainly emphasize the importance of discriminating between ISIS and Islam. The third tweet is coming from a Muslim who condemns the attacks. The fourth tweet wonders why people think ISIS represent Islam, whilst ISIS kills Muslims. The last tweet mocks the media outlets that generalize when an attack is perpetrated by a Muslim or an African American, while they do not generalizing when the attacker is White. 

\subsubsection{Accounts promoting Anti-Islamic rhetoric}
Running a very active campaign in the 2016 presidential race, Donald Trump tweet came first in the list of most retweeted tweets in the attacking category ``Why won't President Obama use the term Islamic Terrorism? Isn't it now, after all of this time and so much death, about time!''. This tweet  
was retweeted 7,050 and received 13,675 likes. Trump had another tweet in the top 5 that was retweeted 2,607 times and received 5,979 likes.  The content of the tweets revolve around the Anti-Islam rhetoric in reference to the Paris Attacks. In the tweets, Trump continues to slam the Democratic Party and president Obama for not referring to the ISIS attacks as “Islamic Terrorism.” When looking at Trumps Timeline, it becomes clear that this is one of many tweets that are following the same sentiment, that is blaming Islam and Muslims worldwide for the Paris attacks, in addition to many other sentiments that serve his political agenda in the presidential race. 

Ted Cruze, another US presidential candidate, also had a tweet in the top 5 with clear linking between Islam and terrorism.
The appearance of yet another tweet from another conservative US politicians may indicate the political nature of the comments and their ties to conservative right-wing mood in the US.

Following Trump’s tweet is a tweet from Ayaan Hirsi Ali, a female human rights activist based in the US with Somali origins, who is known for her critical view on Islam and has been actively calling on public forums for a reformation of Islam~\footnote{\url{https://en.wikipedia.org/wiki/Ayaan_Hirsi_Ali}}. In her tweet Ayaan writes, ``As long as Muslims say IS has nothing to [do] with Islam or talk of Islamophobia they are not ready to reform their faith'' Ranking number 3 on the list with 5,800 retweets and 5,318 likes. Ayaan calls on all Muslims around the world to consider and recognize that Islam as a faith is sourcing terrorist ideology. Ayaan is affiliated with the American Enterprise Institute, a right-wing conservative think tank in the US, which may also indicate yet another link between her comments and US politics.  This tweet was ranked as `attacking' as per our crowdflower results.  

Interestingly, from our high level analysis (going back to the accounts and reading previous tweets) it appears that the list of most retweeted tweets attacking Islam is dominated by controversial figures who have expressed some anti-Muslim sentiment on previous occasions, which helps in predicting this kind  of behavior. However, the relationship between someones social background and their vocalization of Islamophobia is out of the scoop of this paper and will be discussed in the future work. 

\section{Conclusion and Future Work}
\label{sec:conclusion}
In this paper, we analyzed the direct response of Twitter users on the tragic terrorist attacks on Paris on November 13, 2015. We applied quantitative analysis to the tweets that were particularity discussed Muslims and Islam in conjunction with the attacks. Our findings in this paper show that in general the vast majority of tweets were positive towards Islam and Muslims. The ratio of positive to negative tweets varied greatly from one language to another and from one country to another, often between neighboring countries.  Concerning the top hashtags that are linked to English tweets that attack Muslims, most seem to be coming from the US and they may be political in nature.

There is an opportunity to continue expanding on this study to answer questions we had during our research. For example, 1) What is/are the motivation/s of the attackers to relate Islam to terrorism? What is their background? Does it have relation to being Islamophobic? 2) Are the majority defenders of Islam and Muslims in fact Muslims? If not, what is their ideology? 3) Why some western countries were more defending than others? Is it related to the media? or history? 4) What lessons can we learn from this incident to inform policy maker and designers on ways to predict the spread of hate speech online and the best ways to deal with it. In addition, knowing the leading accounts and type of messages being spread online after crises is import to better contain this phenomenon to prevent its consequences from leaving the virtual world.

The utilization of Twitter as platform for political engagement is not recent nor new~\cite{akoglu2014quantifying,wong2013quantifying,liu2014quantifying}, however, the recent trends of hate speech appearing on the network are alarming. We believe that the results of this work will have a positive impact on the research being conducted on adversarial (or perhaps hate) speech online and potential spell over into the non-virtual world. 



\begin{footnotesize}
\bibliographystyle{aaai}
\bibliography{ref}
\end{footnotesize}

\end{document}